\documentclass[aps,prl,showpacs,twocolumn]{revtex4}
\usepackage{graphicx}
\begin{document}
\newcommand {\be}{\begin{equation}}
\newcommand {\ee}{\end{equation}}
\newcommand {\bea}{\begin{eqnarray}}
\newcommand {\eea}{\end{eqnarray}}
\newcommand {\nn}{\nonumber}


\title{ Ground state of a two dimensional quasiperiodic quantum antiferromagnet}

\author{ A. Jagannathan}
\affiliation{Laboratoire de Physique des Solides, Universit\'e Paris-Sud,
91405 Orsay, France \\
}

\date{\today}

\begin{abstract}
We consider the antiferromagnetic spin-1/2 Heisenberg model on a
two-dimensional bipartite quasiperiodic tiling. The broken
symmetry ground state in this model is inhomogeneous, but
nevertheless bears interesting similarities with that of the
square lattice antiferromagnet. An approximate block spin
renormalization scheme developed first for the square lattice is
generalized here to the quasiperiodic case. The ground state
energy and local staggered magnetizations for this system are
calculated, and compared with the results of a Quantum Monte Carlo
calculation for the tiling.

\end{abstract}
\maketitle

\section{\label{sec:level1}I. The experimental and theoretical background}
Magnetism in quasicrystals can be very complex, due to the extreme
sensitivity to structural details, in such systems, of local
moment amplitudes as well as of the interactions. A considerable
simplification of the problem is however possible for the recently
studied rare-earth based quasiperiodic alloy ZnMgHo \cite{sato}.
The rare-earth based magnetic alloys represents a conceptually
simpler system than the transition metal alloy quasicrystals that
were initially the object of experimental studies, since the
magnetic moments are associated with f-orbitals, and can
 be assumed in the first approximation to be {\it independent} of the
local itinerant-electron density of states.

This is to be contrasted with the earliest magnetic quasicrystals
of the AlMn family, where the itinerant magnetic moments on the Mn
 atoms depend sensitively on detailed structural
features due to the d-orbital hybridization (see the review by
Hippert et al in \cite{belin}). To add to the difficulties the
early alloys were metastable quasicrystals of inferior structural
quality so that the role of disorder had to be considered in
addition to the intrinsic behavior. Experimental results indicated
a wide distribution of effective moments on the Mn atoms, as well
as of the interactions between these, leading to a large number of
unknown parameters in the phenomenological models describing such
systems. From a theoretical viewpoint, therefore, the rare earth
system is clearly far simpler.

 ZnMgHo was shown to undergo a magnetic
transition into a magnetic state characterized by short range
antiferromagnetic correlations with quasiperiodic modulation
\cite{sato}. The experimental results lead naturally to the
question of what properties one expects for the ground state of a
quasicrystal with short range antiferromagnetic interactions. An
acceptable starting point for models of such systems could be, as
for crystalline compounds, a Hamiltonian with short range
antiferromagnetic couplings between pairs of identical spins, $H =
\sum J_{ij} {\bf S}_i.{\bf S_j}$.

\begin{figure}[ht]
\begin{center}
\includegraphics[scale=0.50]{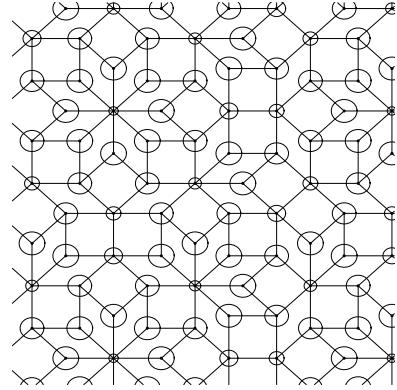}
\vspace{.2cm} \caption{Inhomogeneous ground state structure on the
tiling. The circles have sizes that depend on the strength
 of the local order parameter
}

\label{first.fig}
\end{center}
\end{figure}

Fig.1 shows the results of a recent Monte Carlo study of a
two-dimensional model of quantum spins on a quasiperiodic tiling
\cite{wess}. The circles on the vertices have radii that depend on
the value of the local staggered moment, a quantity that we will
define further below. The tiling considered is the eight-fold
symmetric octagonal (Ammann-Beenker) tiling, in which sites can
have six possible values of coordination number $z$. Sites were
occupied by $S=\frac{1}{2}$ spins, with uniform interactions
$J_{i,j}=J >0$ along the edges of the tiling. The system is
bipartite, meaning that every spin belongs to one of two
sublattices and interactions couple only spins of different
sublattices. Analogously to the spin $\frac{1}{2}$ square lattice
antiferromagnet, which is now believed to have a ground state with
long range order, we expect
 that this quasiperiodic  system, too, has a broken symmetry ground
state with long range order. Classically, the ground state
corresponds to having oppositely directed sublattice
magnetizations, with no frustration, in the sense that all bonds
can be ``satisfied" simultaneously. In the quantum case, the
ground state will correspond to zero total spin since for the
octagonal tiling, the two sublattices are equivalent.

The inhomogeneous structure of the ground state seen in Fig.1 is a
reflection of the environment-dependence of the quantum
fluctuations around the Neel state in the quasicrystal. There is
for the moment no spin wave expansion that would allow, as in a
periodic solid, to calculate the distribution of staggered moments
and explain the QMC results. In fact, as we will see, a real space
approach seems more appropriate for quasiperiodic tilings, and
many such calculations exist for the case of one dimension.

One-dimensional models to study the behavior of quantum spins on
quasiperiodic chains been considered by several authors. Quantum
spin chains have been analyzed using renormalization schemes
\cite{herm,hida} based on the inflation symmetry of these chains.
 Using a mapping to fermionic models and techniques of bosonization,
\cite{vidal}  it is possible to obtain interesting results
concerning global properties such as the magnetization as a
function of external field, and the spectral gaps for a variety of
different quasiperiodic sequences. However, real space information
such as the distribution of the local quantities $m^2_{loc,i} \sim
\vert\langle S_i S_{i+1}\rangle\vert$ has not so far been
calculated.

For two dimensional structures, real space configurations have
been studied for models with classical spins. Here, the ground
state is nontrivial only when the model includes frustration. In
\cite{lucgod} Godreche et al introduced a renormalization scheme
on the Penrose tiling for a Heisenberg exchange model with
competing antiferromagnetic interactions, and were thus able to
obtain a phase diagram consisting of a variety of ordered phases.
The real space spin configurations were calculated numerically in
\cite{ved} for classical spins interacting via long-ranged dipolar
interactions, and a complex magnetization distribution
 with overlapping decagonal rings
reflecting the underlying Penrose tiling was found. A
quasiperiodic magnetic state with a heirarchical ringed structure
was found, as well, in a different context: that of itinerant
magnetism due to interacting electrons \cite{jag2}.

 With this background, we return to the problem of quantum
 Heisenberg spins with nearest neighbor antiferromagnetic interactions.
 Ref. \cite{wess} presented
   local staggered order parameters similar to the quantities $m_{loc,i}$
 above, calculated
for individual sites using the expectation values for local
spin-spin correlations. One sees in Fig. 1 that sites of the same
$z$ have similar local order parameter amplitudes. An explanation
of this behavior was given by considering isolated star shaped
clusters called Heisenberg stars in \cite{wess}. This provided a
qualitative understanding of the decrease of local staggered
magnetizations as a function of $z$, but for a more quantitative
fit to the QMC results, it is necessary to go beyond the isolated
cluster approximation, and take into account longer range
correlations. This can be done in a renormalization group (RG)
calculation that uses
 an important symmetry of the tiling, namely
invariance under discrete scale transformations called inflations.
 This renormalization group is a generalization of the
calculation of Sierra and Martin-Delgado for the square lattice
\cite{sierra}, where the authors considered star-shaped block
spins formed by a central spin and its four nearest neighbors. In
their calculation, block spins formed from these five-spin
clusters are shown to interact via an effective Heisenberg
antiferromagnetic interaction on a bigger $\sqrt{5} \times
\sqrt{5}$ square lattice. The effective spin values scale to
infinity, i.e. the classical limit, under renormalization. Their
model for a translationally invariant system can, as we will see,
be adapted to our quasiperiodic case under certain approximations.
We thus calculate not only the global ground state energy as was
done for the square lattice, but also the distribution of local
order parameters. We will discuss the method, which has been
briefly reported in \cite{jag}, in some detail in the present
paper.

We begin with an introduction to the quasiperiodic tiling and the
spin Hamiltonian in the next two sections. The RG scheme is
described in the fourth section. Results and discussions are
presented in sections five and six.

\section{\label{sec:level1}II. Review of geometrical aspects}

\begin{figure}[ht]
\begin{center}
\includegraphics[scale=0.40]{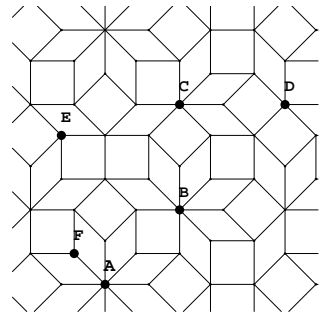}
\vspace{.2cm} \caption{A portion of the octagonal tiling showing
the six different nearest neighbor environments A,B,...,F}
\label{tiling.fig}
\end{center}
\end{figure}
\subsection{\label{sec:level2}1. Some general remarks}

The octagonal tiling \cite{soco} shown in Fig.\ref{tiling.fig} can
be thought of as the equivalent of the square lattice for
quasiperiodic systems. It has therefore been frequently used for
analytical and numerical investigations of the effects of
quasiperiodic modulations in two dimensions. Spectral properties
of electrons \cite{sire}, transport properties \cite{trans},
 vibrational properties \cite{janss} and magnetic
properties \cite{wess} have thus been studied for discrete models
defined on the octagonal tiling. The tiling is built from two
kinds of tiles, squares and $45^o$ rhombuses. These two types of
tiles can fill the two-dimensional plane in an aperiodic way, as
Penrose first showed for the five-fold tiling named after him
\cite{pen, gard}.

 Although there is no translational
invariance in a quasiperiodic tiling, any given tile arrangement
of tiles reoccurs all over the tiling with a certain frequency of
re-occurrence  -- or, alternatively viewed, there exists a mean
distance of separation between such identical domains. This is
referred to as the repetitivity property of quasiperiodic tilings,
and is very different from the situation in a disordered medium
(where the expected distance in which one expects to find a second
region identical to the first increases exponentially with the
size of the region). Similarly, the property of symmetry under
rotations for these tilings differs from that in crystals, for
which the new and the old structures coincide exactly. For the
quasicrystal, the equivalence of the new and old tilings holds in
the ''weak" sense, namely, any finite region of the new tiling
after rotation will be identical to finite regions of the old one.

\begin{figure}[ht]
\begin{center}
\includegraphics[scale=0.70]{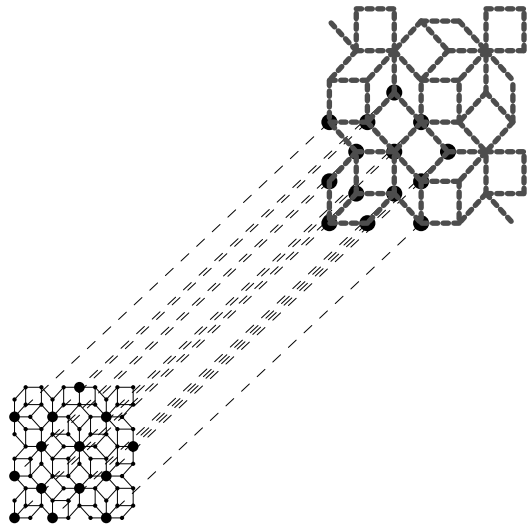}
\vspace{.2cm} \caption{Portion of original (black) tiling, showing
sites of the $\alpha$ class (black dots) which become sites of the
new inflated (grey) tiling} \label{infl.fig}
\end{center}
\end{figure}

Such aperiodic structures can be built using "matching rules".
These are local rules that determine if and how two tiles can be
laid side by side (see Ch.1 of \cite{stein}). Alternatively,
tilings such as the Penrose and octagonal tilings could be
generated by a projection method down from a higher dimensional
periodic structure \cite{proj}. Such an approach can give either a
deterministic, perfectly ordered tiling, or a random one where
tiles are assembled subject only to the constraint that they
should fill space without overlapping \cite{hen}. Random tilings
are of great theoretical interest,
 but we are here interested in
deterministic tilings, which have the important property of
invariance under inflation/deflations, or discrete scale
invariance. This symmetry is illustrated in Fig. \ref{infl.fig}
and will be described in more detail in the next section. It is
this property that is responsible for the characteristic singular
electronic and magnetic properties of such tilings and it
 was first pointed out in the Penrose
 tiling, which is invariant under a replacement of tiles by $\tau$-fold bigger
tiles, where $\tau = (\sqrt{5}+1)/2$ \cite{gard}. One can define
geometrical inflation rules for, among others, the Fibonacci chain
in one dimension, the octagonal tiling in two dimensions, and the
icosahedral tiling in three dimensions.

The renormalization approach is a natural one for such
geometrically self-similar quasiperiodic tilings , and this
structural property has been exploited in order to establish
recurrence relations for parameters occurring in discrete spin
models, electron hopping models, etc, as mentioned before for the
one-dimensional case, but also for some two-dimensional models
\cite{lucgod,moss2}, where analytical methods remain hard to
implement. As noted in the introduction, our approach is inspired
by the renormalization calculation of Sierra and Martin-Delgado
\cite{sierra} for the square lattice.

 Some principal properties of the octagonal tiling that are used
 in the RG calculation are reviewed in the next section,
  without demonstration. (For those
interested, Appendix A contains some additional details on how to
obtain a quasiperiodic structure, and how inflations/deflations
are described in the framework of the projection method. Although
not strictly necessary to understand the calculations presented
below, an understanding of the geometrical properties of the
tiling is important for those wishing to improve this approximate
RG scheme and extend it to other models. For more details, the
reader is referred to reviews in \cite{lucrev,dunref}).

\subsection{\label{sec:level2}2. The six local environments}

\begin{figure}[ht]
\begin{center}
\includegraphics[scale=.90]{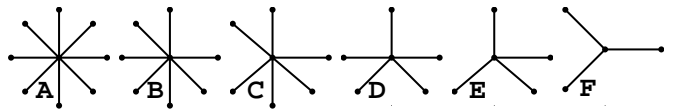}
\vspace{.2cm} \caption{ The six different nearest neighbor
environments of the octagonal tiling } \label{sec.fig}
\end{center}
\end{figure}

The six nearest neighbor configurations,
 corresponding to coordination numbers
$z=8,7,...,3$ are labeled  A,B,...,F as shown in Fig.2. Fig.
\ref{sec.fig} shows these environments separately. In an infinite
tiling, each of these types of site occurs with a well-defined
frequency $f_i$, where (see Appendix A)

\bea
 f_A=\lambda^{-4};
 f_B=\lambda^{-5}; f_C=2\lambda^{-4}; \nonumber \\
f_{D1}=\lambda^{-3}= f_{D2} ; f_E = 2\lambda^{-2}; f_F =
\lambda^{-2} \eea with $\lambda=1+\sqrt{2}$. One distinguishes
between two kinds of D sites as explained in the next section. It
can be checked using the above frequencies that the average site
coordination number on the octagonal tiling is exactly four.

\subsection{\label{sec:level2}3. The inflation transformation}
Inflation proceeds as follows for the octagonal tiling: one starts
with a tiling composed of tiles of a given initial edge length (we
will assume this is equal to 1) and one reconnects a precisely
determined subset of vertices so as to obtain a new tiling of the
same type as the old, i.e. having the same set of local
geometries, except for an overall scale change by a numerical
factor $\lambda = 1+\sqrt{2}$ ( Fig.4). The sites shown as black
dots on the original tiling belong in the $\alpha$ class: A,B,C
and half the D (called $D_1$) sites. These become the sites of the
new bigger tiling, while the remaining ($\beta$) sites drop out.
Note that there are two varieties of five-fold sites, $D_1$ and
$D_2$, which belong to the $\alpha$ and $\beta$ classes
respectively. On the octagonal tiling, they always occur in pairs.
Appendix A shows how the two classes of D sites can be
distinguished in terms of their perpendicular space coordinates
\cite{note2}.

Under inflation, the density of sites is reduced to $1/\lambda^2$
of its initial value.  The sites that remain acquire new values of
the site coordination numbers ${z'} \leq z$. The table below lists
the initial and final values of coordination number for each of
the $\alpha$ class sites (note that there are four different
subcategories for the A sites -- see Appendix A for more on the
properties of these subcategories).

\begin{center}
\begin{tabular}{|l c r|}
\hline
 initial site && final site \\
 (z value ) && (z value) \\
\hline
A (8) &\quad \quad $\rightarrow$ \quad \quad &A,B,C or $D_1$ (8,7,6,5) \\
B (7) &\quad \quad $\rightarrow$ \quad \quad & $D_2$ (5) \\
C (6) &\quad \quad $\rightarrow$ \quad \quad & E (4) \\
$D_1$ (5)&\quad \quad $\rightarrow$ \quad \quad & F (3)\\
\hline
\end{tabular}
\end{center}

\noindent {\small Table 1. List of $\alpha$ sites and their
transformations under inflation}

\subsection{\label{sec:level2}4. Nearest neighbors of $\alpha$ sites}

For the four types of $\alpha$ sites, the table below lists the
nearest neighbors (nn) in terms of the type of site and the number
of sites of that type. This information will be useful in
determining the final block spin value at the central site, as we
will explain in sec.III.

\begin{center}
\begin{tabular}{|l|r|}
\hline
$\alpha$ site &  nn spin type(number)   \\
\hline
A& F (eight) \\
B & F (five), E (two) \\
C & F (two), E (four) \\
$D_1$& $D_2$ (one),E (four) \\ \hline
\end{tabular}
\vskip 0.5cm \small{Table 2. The $\alpha$ sites and their nearest
neighbor environments}
\end{center}

Table 2, in conjunction with Table 1 allows one to deduce how
blocks are organized in the tiling. An A site which transforms to
an A site after inflation corresponds, on the original tiling to
an A block surrounded by eight $D_1$ blocks.

\section{\label{sec:level1}III.  The spin Hamiltonian}

We consider onsite spins $\bf{S}_i$ ( i=1,N) where all spins have
spin $\frac{1}{2}$, with the Hamiltonian $H(N,\{S_i\},\{J_{ij}\})$

\be H = \sum_{\langle i,j\rangle} J_{ij} {\bf S}_i.{\bf S}_j \ee

 where $\langle i,j\rangle$ denotes a pair of spins
$\bf{S}_i$ and $\bf{S}_j$  linked by an edge, and $J_{ij}=J>0$ for
such a pair. This system is bipartite with two identical (to be
understood in the weak sense) subtilings (as on the square
lattice).

\subsection{\label{sec:level2} 1. Finite spin clusters}
\begin{figure}[ht]
\begin{center}
\includegraphics[scale=0.8]{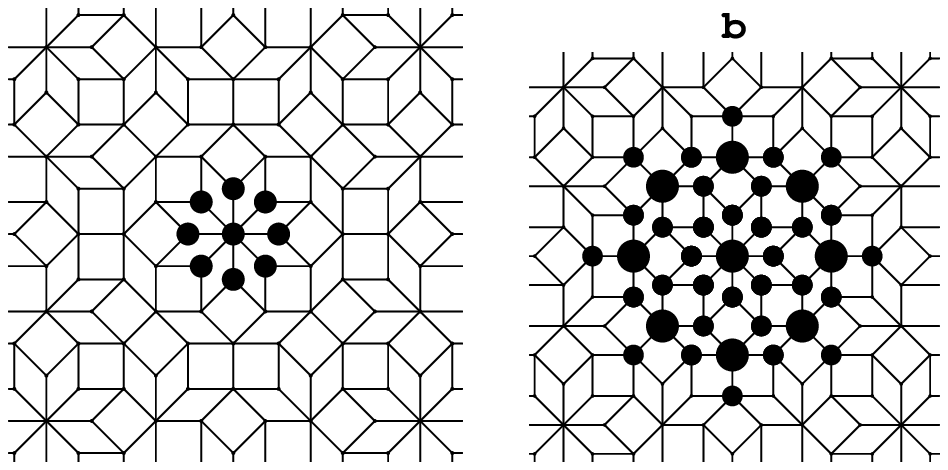}

\includegraphics[scale=0.8]{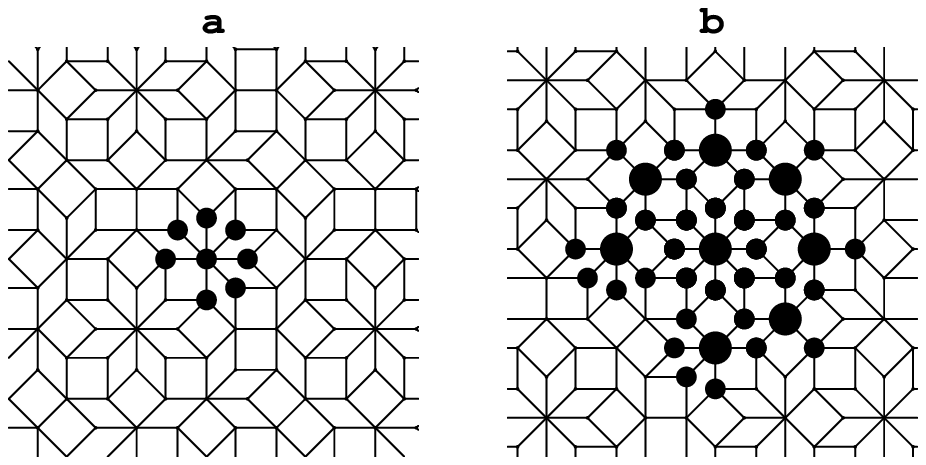}

\includegraphics[scale=0.8]{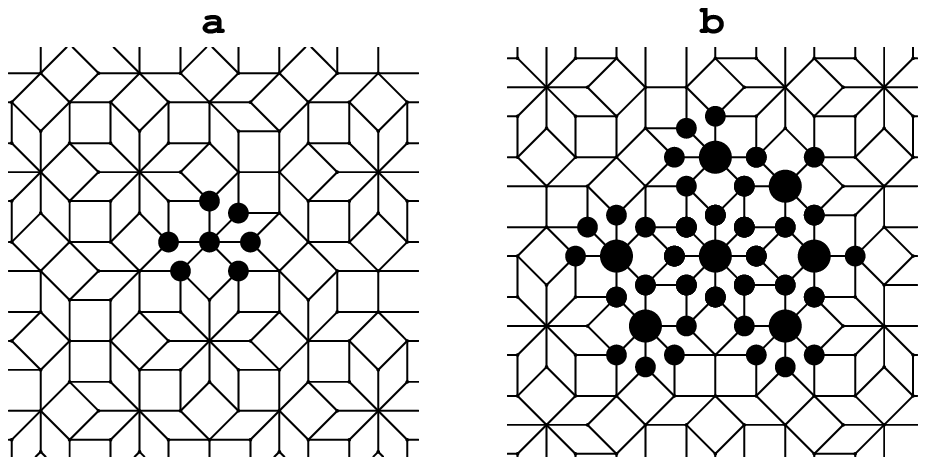}

\includegraphics[scale=0.8]{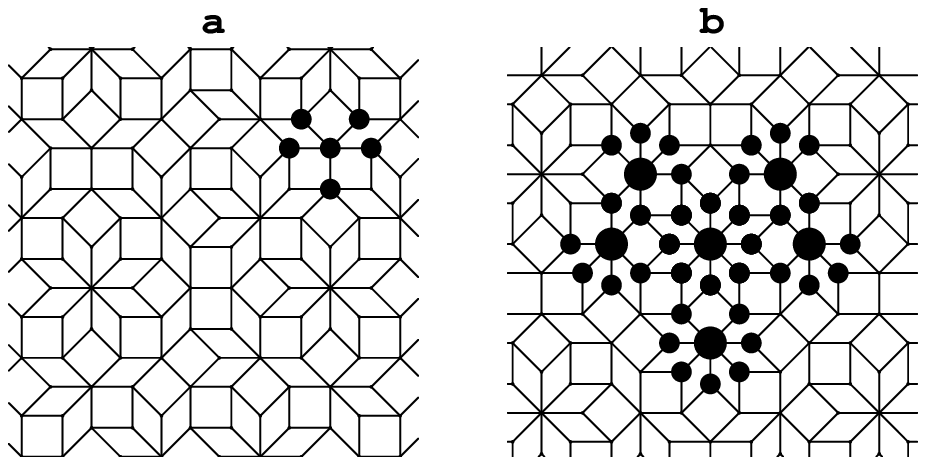}
\vspace{.1cm} \caption{ The $\alpha$ site clusters defined on the
original (left) and once inflated (right) tilings}
\label{clus.fig}
\end{center}
\end{figure}

 A sites are surrounded by eight F sites. If one isolated one such
cluster of 8+1 spins, the lowest energy state for classical spins
is the one with the eight peripheral spins antiparallel to the
central spin. In the quantum case, the ground state of the cluster
is rotationally invariant, and corresponds to the total cluster
spin value $S_{tot} = 7/2$. The other $\alpha$ sites correspond to
total cluster spin values in the ground state of $S_{tot} = (z_B
-1)/2 = 3$ around a B site,  and so on. The four clusters are
shown in the left hand side series of Fig.\ref{clus.fig}.

\begin{figure}[ht]
\begin{center}
\includegraphics[scale=0.8]{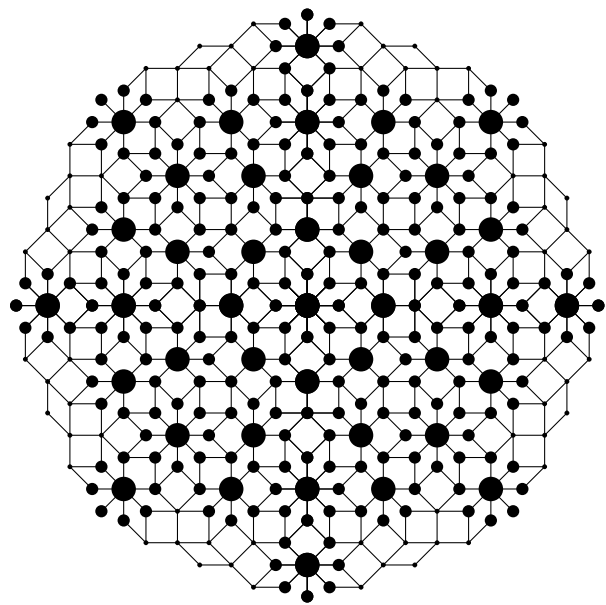}
\vspace{.1cm} \caption{ Second generation A cluster }
\label{alpha2.fig}
\end{center}
\end{figure}

Clusters of each type can be defined on larger and larger length
scales, by using the inflation rules already outlined to determine
the new A,B,C and $D_1$ sites after inflation. Fig.\ref{clus.fig}
shows the four $\alpha$ clusters on the next largest length scale
on the right hand side series. Here, block spin centers are shown
with big black dots, while the sites corresponding to the $\beta$
sites are indicated by smaller dots. On a yet bigger length scale,
Fig.\ref{alpha2.fig} shows a ``second generation A site", namely,
a site that remains of A type after two inflations, along with all
the sites belonging to the cluster before the two decimations.

The $\alpha$ clusters on all length scales are the building blocks
for the renormalization scheme that follows.

\section{\label{sec:level1}IV. The renormalization
transformation} The renormalization calculation is a
generalization to an aperiodic system of the one used for the
square lattice by Sierra and Martin-Delgado \cite{sierra}. We
review briefly the steps of their calculation before showing how
they are modified in the quasiperiodic case.

\subsection{\label{sec:level2} 1. RG on the square lattice}
\begin{figure}[ht]
\begin{center}
\includegraphics[scale=0.450]{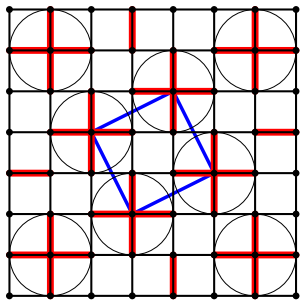}
\vspace{.1cm} \caption{Five-spin units (surrounded by circles) on
the square lattice. The new $\sqrt{5} \times \sqrt{5}$ unit cell
is shown} \label{sqlatt.fig}
\end{center}
\end{figure}

  We consider the nearest neighbor Heisenberg
  antiferromagnet described by Eq. 2 with spin $\frac{1}{2}$ on the vertices and the
  initial coupling $J$ along the edges of the squares
  (of side $a=1$). Fig.\ref{sqlatt.fig} shows the five-spin blocks
  enclosed by circles. The four couplings inside each block
  are shown outlined by thick grey lines. As one sees, the block
  spins form a new rotated square lattice
   of side $\sqrt{5}$ (Fig.\ref{sqlatt.fig}). Each of the blocks
   can be
    diagonalized exactly. With every step of RG, only the lowest energy
    states of the blocks are retained to form the basis for the effective
    Hamiltonian.  $T_0$ and $T^\dag_0$ denote the operators
     describing the
    transformations from the original Hamiltonian (acting in the
    complete Hilbert space) to the effective Hamiltonian (acting
    in the reduced Hilbert space). For a single block, the lowest energy
    sector corresponds to spin $\frac{3}{2}$, and
 the ground
state energy is $e_0 = -JS(4S+1)$. The couplings not already taken
into account give rise to inter-block interactions, calculated by
first order perturbation theory. It is easy to check that the new
block spins will be coupled antiferromagnetically to its nearest
neighbors, like the original spins. The effective Hamiltonian
$H(N,S,J)$ can thus be written approximately as a sum of
single-block contributions (the diagonal terms) and a set of terms
involving nearest neighbor blocks (off-diagonal terms), and
 the formal expression for the transformed problem reads

\begin{eqnarray}
T^\dagger_0 H(N,S,J) T_0 = N'e_0(J,S) + H'(N',S',J') \nonumber \\
\end{eqnarray}

where the new Hamiltonian $H'$ has the same form (bilinear in
$S'$) as $H$, and $N' = N/5$. The effective spin of a block spin
is $S'= 3S =\frac{3}{2}$ . The spin renormalization factor
relating one of the four boundary spins to the new block spin has
been shown to be close to the classical value $\xi_0 = S_i/S'
\approx \frac{1}{3}$ ( see \cite{sierra} for the exact value). The
interaction between two contiguous blocks is $J' = 3\xi^2_0 J$.

Repeating the steps of renormalization, one has ultimately for the
ground state energy per site an infinite sum as follows

\begin{eqnarray}
e_\infty = -\frac{1}{5} \sum_{n=0}^\infty 5^{-n} J^{(n)} S^{(n)}
(4 \times 3^n S +1)
\end{eqnarray}

where $S^{(n+1)} = 3 S^{(n)}$ and $J^{(n+1)}= 3\xi^2(S^{(n)})
J^{(n)}$. Under RG, the spins evolve to the classical limit, $S
\rightarrow \infty$ indicating that in the quantum case as well
one has a ground state with broken symmetry. The couplings scale
to zero indicating the model is massless. Qualitatively, thus, the
RG gives the now accepted physics of the model, however,
quantitatively the value obtained for $e_\infty \approx -0.546$ is
not as good as that obtained by spin wave expansion and is about
15 \% higher than that established by numerical calculations
\cite{sand}. We will return to this point at the end of the paper.

\subsection{\label{sec:level2} 2. RG on the octagonal tiling}
On the octagonal tiling, it is clear that several kinds of block
spins must be introduced. A natural choice is to designate the
$\alpha$ sites as block spin centers. Fig.\ref{four.fig} shows the
positions of the block spins (black dots) on a portion of the
tiling. Upon inflation, the other sites will disappear, leaving
only the block variables, and some residual interactions between
them. If no new couplings are generated, one will find an
effective Hamiltonian similar to the old, except for the
renormalized couplings which become site dependent. One can repeat
the process, and determine if there is convergence to a fixed
point.

The simple scheme outlined above cannot be implemented without
some modifications and approximations. The first problem arises
because the connectivity of the tiling is such that some of the
block spins overlap, that is, share two intermediate $\beta$ sites
in common. This is shown by the thick grey lines in
Fig.\ref{four.fig}, which indicate the boundary between
overlapping blocks. Overlapping occurs between contiguous $C$ and
$D_1$ blocks, as well as between contiguous $D_1$ blocks. This
overlapping occurs with a finite density. One can calculate this
density by noting that the shared sites occur between any two
sites that are a distance $\lambda^2 d_s$ apart, where $d_s$ is
the short diagonal of the rhombus. One finds, using the relative
frequencies of occurrence of squares and rhombuses that the
density of pairs is $\sqrt{2}/\lambda^3 $, that is, about 10\% of
the total number of pairs.

\begin{figure}[ht]
\begin{center}
\includegraphics[scale=0.60]{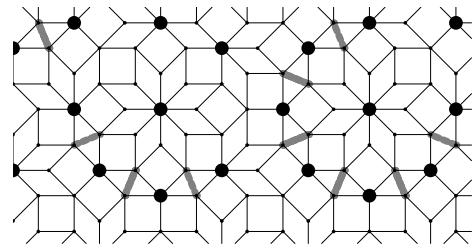}
\vspace{.2cm} \caption{Tiling showing block centers (black dots).
The grey lines connect pairs of sites that are shared between two
blocks. } \label{four.fig}
\end{center}
\end{figure}

 To deal with this problem, we therefore considered two possible
modifications of the original model, i) doubling the number of
spins on each shared site, and considering each spin as being
coupled to one block only, and ii) decoupling the block spins by
annulling one of the bonds to the left or the right so that spins
are no longer coupled on both sides. The first modification leads
to overestimating the total energy, the second to underestimating
it, with respect to the original octagonal tiling. Spin doubling
on selected sites leads to an uninteresting flow under
renormalization, where cluster energies basically repeat a scaled
Heisenberg star distribution at each step. The bond dilution
scheme yields a more complicated behavior of cluster energies
under renormalization, and is the option taken up in detail in
this paper.

We note that the diluted model remains two-dimensional, and is not
of a scale invariant fractal such as the Sierpinski gasket
\cite{havlin}, where bonds are also deleted heirarchically but in
a way that leads to an effective
 fractal dimension less than two.

The second problem is the quasiperiodic connectivity between
blocks which leads ultimately to an infinite number of
environments. This is dealt with by truncating the number of
environments we choose to distinguish between. The $\alpha$ sites
always have the same type of nearest neighbors (given in Table 2),
however the $\beta$ sites occur in several configurations. We will
now truncate the table of connectivities by allowing only one type
of $D_2$,E and F site, and a connectivity table as follows:

\begin{center}
\begin{tabular}{|l|r|}
\hline
$\beta$ site &  nn spin type(number)   \\
\hline
$D_2$&  $D_1$ (one),E (two),F(two) \\
E & $\alpha$ (two), F (two) \\
F & $\alpha$(one), E (two) \\
 \hline
\end{tabular}
\vskip 0.5cm \small{Table 3. The $\beta$ sites and the truncated
set of nearest neighbor environments}
\end{center}

\subsubsection{\label{sec:level3}A. Bond dilution and the new block
spins}

\begin{figure}[ht]
\begin{center}
\includegraphics[scale=0.450]{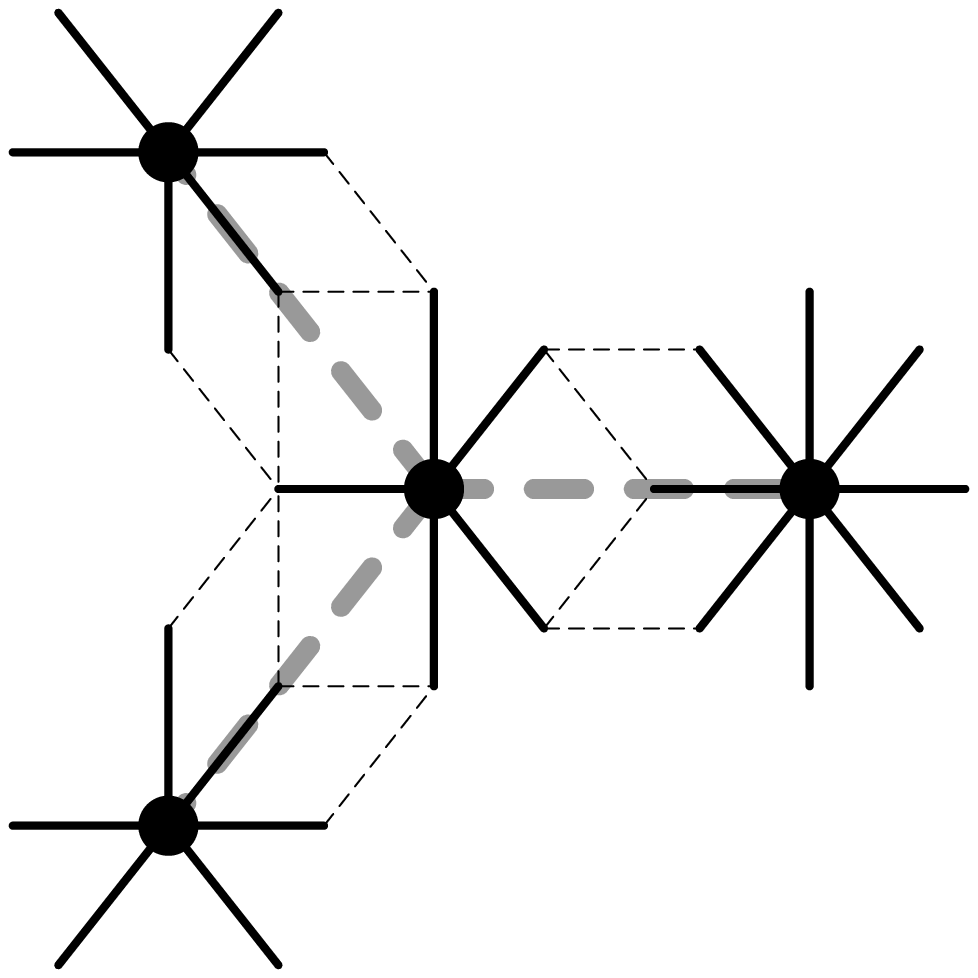}
\includegraphics[scale=0.450]{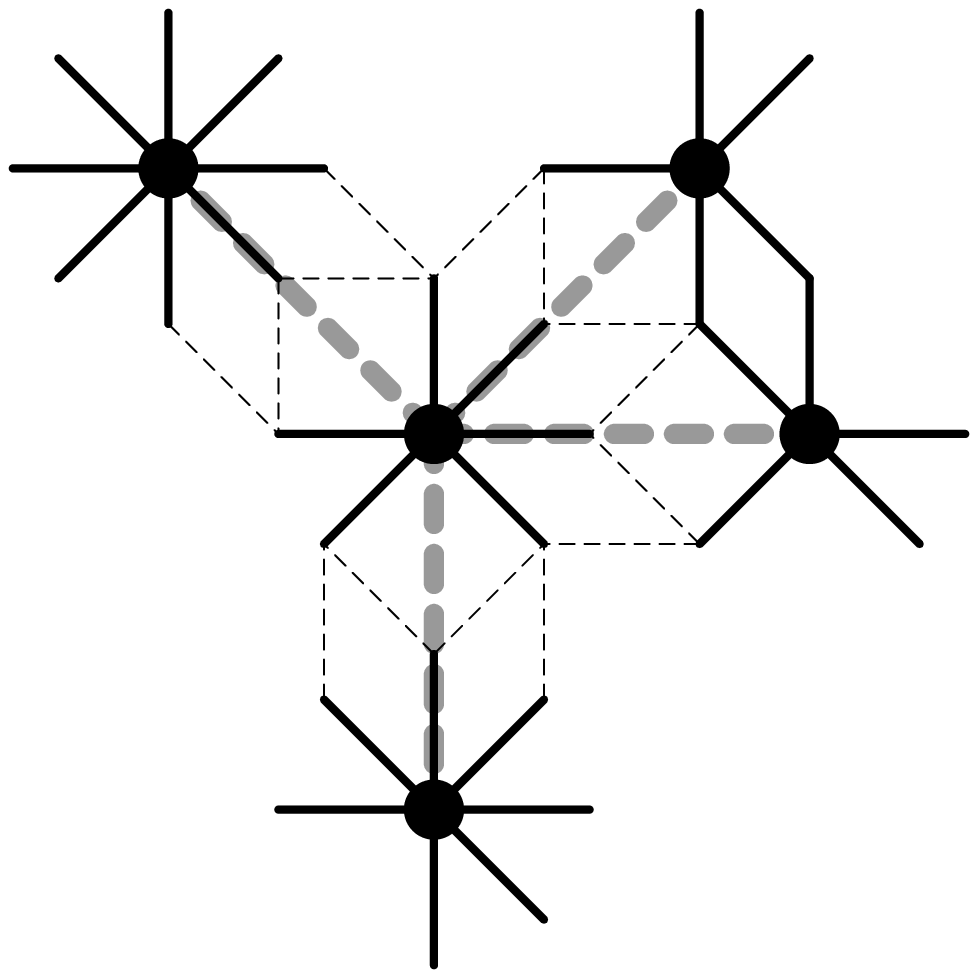}
\includegraphics[scale=0.450]{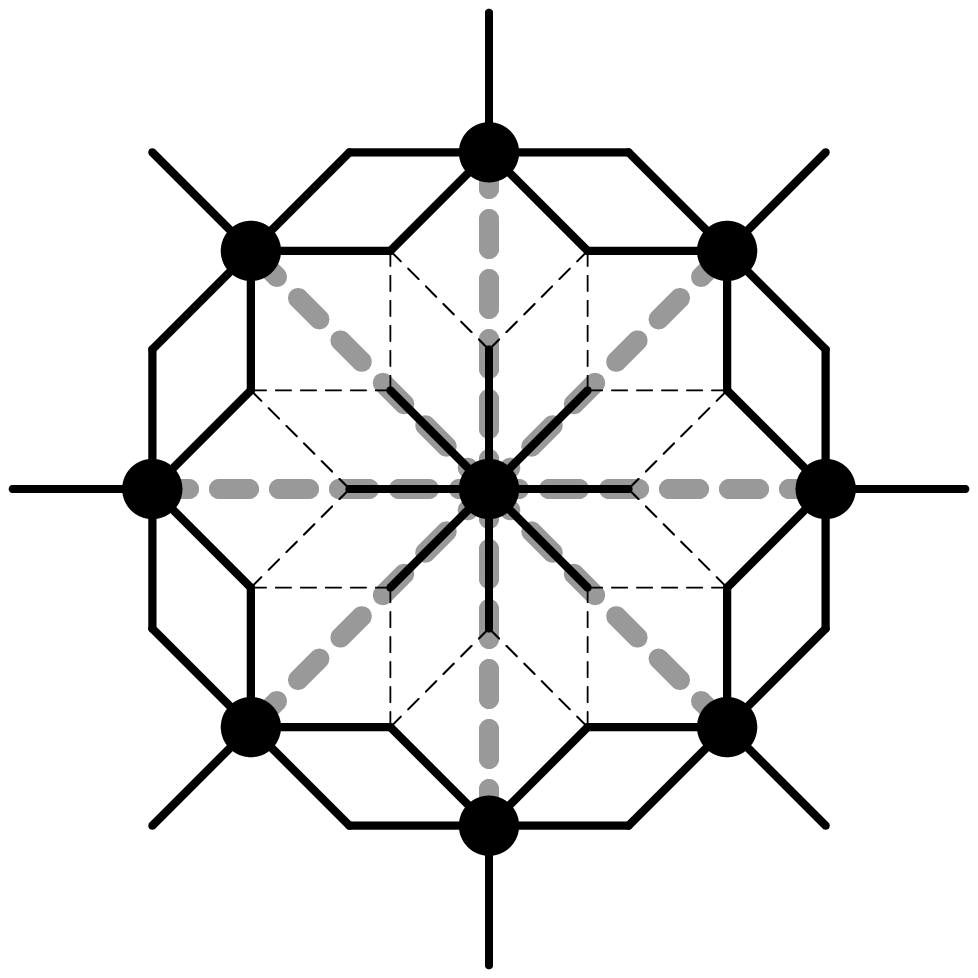}
\vspace{.1cm} \caption{Block spin centers (filled circles) showing
the central and all peripheral blocks for three cases:
 (top) a $z=5, z'=3$ site
(middle) a $z=6, z'=4$ site  (bottom) a $z= z'= 8$ site }
\label{sixth.fig}
\end{center}
\end{figure}

In this subsection we discuss the blocks that are obtained after
dilution and the values of the effective block spin.
Fig.\ref{sixth.fig}(top) shows in detail a central $D_1$ site
which transforms to an F site under inflation. The three
neighboring block spins are shown as well, with the block spin
sites shown by black dots. The original links are indicated by
thin black lines, while the new effective links on the inflated
tiling are shown by thick grey lines. Fig.\ref{sixth.fig}(middle)
shows a C site transforming to an E site, with the same
conventions used to denote block spin sites and new effective
couplings. In this figure one sees that two of the block spins,
corresponding to neighboring $D_1$ blocks, overlap. The pair of
sites shared between the two blocks is coupled to the left and
right by a total of four bonds. In the bond-dilution approach, one
has to set two of the bonds equal to zero. This can be done in one
of two ways that treat the two blocks equitably, leaving each
$D_1$ block with one less bond. Finally,
Fig.\ref{sixth.fig}(bottom) shows an A site transformed under
inflation to a final A site. In this case, the eight $D_1$ blocks
surrounding the center block form a ring of overlapping blocks.
There are two ways to decouple them all by annulling eight of the
sixteen links joining them in way that treats all the $D_1$ sites
equitably. Ultimately, the bond dilution results in an effective
reduction of connectivity of C and $D_1$ sites: the former have
the effective $z$ value $\tilde{z} = 5$ and the latter $\tilde{z}
= 3$.

\subsubsection{\label{sec:level3}B. Spin renormalization factors}

Consider a block spin composed from a cluster of $z$ spins
surrounding a central spin and antiferromagnetic interactions. In
the simplest case where all spins have the value S, the block has
a spin of $S' = (z-1)S$ in the ground state. The
 spin
renormalization factors are taken to be equal to the classical
value for simplicity, so that for a given block $\xi_z
=(z-1)^{-1}$.
 The new block spins $S'$ are
situated on the black circles representing the sites of the
inflated lattice, while all of the nearest neighbors are decimated
in the renormalization group (RG) transformation. Initially, all
spins have the same value of spin, $s_0 = \frac{1}{2}$, so that
after one inflation the block spin variables are simply ${\bf
S}^{(1)} = \{S^{(1)}_A,.....,S^{(1)}_F\} =
\{7s_0,7s_0,7s_0,7s_0,6s_0,4s_0,2s_0\}$. (Note that for $E$ and
$F$ sites, the value of z was corrected for the bond dilution). In
subsequent inflations, one has the following matrix  relation
 ${\bf{S}}^{(n)} =
(S^{(n)}_A,...,S^{(n)}_F) = C {\bf{S}}^{(n-1)}$, with

\bea C = \left(
\begin{array}{rrrrrrr}
-1&0&0&0&0&0&8\\
-1&0&0&0&0&0&8\\
-1&0&0&0&0&0&8\\
-1&0&0&0&0&0&8\\
0&-1&0&0&0&2&5\\
0&0&-1&0&0&3&2\\
0&0&0&-1&1&2&0\\
\end{array}
\right) \eea Note that the number of values of the block spins
after each inflation does not grow -- there are just four possible
different values of the block spin at any stage of inflation. As
in the square lattice example, the spins all tend to the classical
limit as $n$ goes to infinity. In addition, the largest eigenvalue
of $C$, 3, is precisely that of the square lattice in section IV.1
! This eigenvalue,
 along with the corresponding eigenvector gives the flow of effective spin values
 in the limit of large $n$. Thus
 for large $n$ ${\bf{S}}^{(n)} \approx 3{\bf{S}}^{(n-1)}$. This
is the same spin renormalization as that on the square lattice
where $z$ is everywhere equal to 4. In both cases, the spins tend
to infinity, i.e. the classical limit, under renormalization. On
the tiling, moreover, the block spins tend to constant relative
asymptotic values which are site dependent and given by the
eigenvector
 $(1,1,1,1,1,\frac{3}{4},\frac{1}{2})$.

\subsubsection{\label{sec:level3} C. Ground state energy of an
isolated block}

\begin{figure}[ht]
\begin{center}
\includegraphics[width=3.0cm,angle=0]{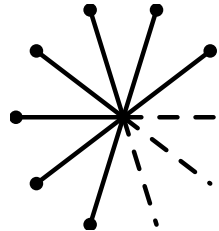}
\vspace{.1cm} \caption{ ($z$+1) spin cluster (Heisenberg Star) }
\label{heis.fig}
\end{center}
\end{figure}

Consider the configuration of $z+1$ spins of Fig.\ref{heis.fig} in
which each of the $z$ links represent the same antiferromagnetic
coupling $J$, termed the Heisenberg Star (HS) in \cite{wess}. For
spin $\frac{1}{2}$ variables on each site and for a given
antiferromagnetic coupling J between the central spin and its $z$
neighbors, the ground state energy can be found exactly to be
\begin{equation}
\epsilon^{(0)}(z) = -J(z + 2)/4 \label{hs}
\end{equation}
On the octagonal tiling, one has the seven different families of
star clusters on the tiling, with the corresponding values of $z$
on the right hand side of the equation. The superscript ''0"
indicates that this corresponds to the energy of unrenormalized
clusters.
 We also require the ground state energy in the case
 of  clusters of  spins of unequal
lengths. The lowest energy state of a cluster in which $z$
 spins of unequal lengths $S_i = n_i s_0$  are coupled with
 strength J
 to a central spin $S_0 =
n_0 s_0$ is taken to be the following generalisation of
Eq.\ref{hs}
\begin{equation}
\epsilon({J},z,\{n\}) = -n_0{{J}}(\sum_{i=1}^z n_i + 2)/4
\label{genhs}
\end{equation}
 In the present model, although initially
the couplings are all equal,
 after one RG step the couplings take on different values. Therefore
we shall make an approximation later that consists of
 replacing the set of couplings around each site by a single
 locally averaged value.

 \vskip 0.5cm

\subsubsection{\label{sec:level3}D. Proliferation of blocks under
deflation}

If ${\bf{n}}=(n_A,n_B,n_C,n_{D1})$ are the number of blocks in a
given region of each given type, the number of blocks of each type
after one deflation is $P{\bf{n}}$ where \be P = \left(
\begin{array}{rrrr}
1&0&0&8 \\
1&0&2&5\\
1&0&4&2\\
1&1&4&0
\end{array}
\right) \label{prolif}\ee The largest eigenvalue of the
proliferation matrix $P$ is equal to 7 so that the total number of
blocks increases(decreases) with the number $m$ of
deflations(inflations) as $7^m$ for large $m$. Notice that the
proliferation of blocks is described by an integer, and not the
irrational number $\lambda^2 \approx 6.8$, each of these numbers
being the answer to a different question. The former describes the
rate of growth of a finite system in terms of the number of
blocks. The latter is the scale factor of the change of site
density under inflation/deflation for the infinite quasicrystal,
and this is not restricted to have integer values.

\subsubsection{\label{sec:level3} E. Renormalization of links}

There are an infinite number of types of links since each link
couples two sites that are each unique. However, just as we chose
to truncate the size of the space of solutions by distinguishing
only seven types of sites, we can consider a ``minimal" model
where it suffices to take into account only five kinds of links.
These are represented in an array ${\bf{j}}=(j_{\alpha
F}$,$j_{\alpha E}$,$j_{D_1 D_2}$, $j_{D_2F}$,$j_{EF})$. Here,
$j_{\alpha F}$ is used to denote the link between (A,F),
(B,F),(C,F) and ($D_1$,F) pairs. Similarly, $j_{\alpha E}$ denotes
the link connecting (B,E),(C,E) and ($D_1$,E) pairs. This
oversimplification of the link classification ignores, in
particular, that E and F sites can occur in more than one
environment. However, in the first approximation, we have assumed
here that one can treat all the sites of a given family as
identical out to first neighbors, and this approximation will be
found {\it {post facto}} to yield reasonably good numerical
results.

 Note that there are no bonds linking sites
that are separated by a distance $d_s$ in the original tiling
(recall that this is the shortest distance possible on the
octagonal tiling) and the same is true for the sites of the
inflated tiling since our bond dilution has the effect of
decoupling such blocks.

Interblock links are all the links not taken into account in the
definition of blocks. To find the new effective links, one also
allows for bond moving, as illustrated by the following example:
consider a central A site surrounded by eight $D_1$-clusters.
These transform to an A site with eight F sites around it after an
inflation. We wish to obtain the effective link between the
central A and one of the F sites. The original A site has sixteen
links to the eight $D_1$-clusters - i.e. it has two links per
$D_1$-cluster. These two links between the center and each
peripheral block are of the $EF$ type (see Fig.8c). Thus the new
effective coupling between the central $A \rightarrow A$ site and
each of the eight $D_1 \rightarrow F$ sites around it on the
inflated tiling is of the $\alpha F $ type. It is
antiferromagnetic, like the original couplings. One takes into
account the spin renormalization factors of the block spins
mentioned before, namely $\xi_A$ and $\xi_D$ respectively. The new
coupling can then
 be expressed in terms of the previous generation of couplings by
the equation

\be j^{(1)}_{\alpha F} = 2 j^{(0)}_{EF} \xi^{(0)}_A \xi^{(0)}_D
\ee

For the second type of links, $j_{\alpha E}$ , one sees that there
are three $EF$ links joining a $A \rightarrow B$ site to a $C
\rightarrow E$ site, so that the new $\alpha E$ link is given by
 \be j^{(1)}_{\alpha E} = 3 j^{(0)}_{EF}
\xi^{(0)}_A \xi^{(0)}_C \ee The other effective couplings can be
written down similarly, although a problem arises due to the fact
already mentioned, namely, that E and F sites can occur in more
than one local environment. Here we chose just one option among
the several, to write down the new effective  $D_2F$ and $EF$
couplings. With this truncation of the link relations, we have a
system of equations between the five old and five new couplings,
${\bf{j}}^{(1)} = M^{(0)} {\bf{j}}^{(0)}$, where
\begin{equation}
M^{(n)} = \left(\begin{array}{ccccc}
0&0&0&0&2 \xi^{(n)}_A \xi^{(n)}_D \\
0&0&0&0&3\xi^{(n)}_A\xi^{(n)}_C\\
0&0&0&0&4\xi^{(n)}_A \xi^{(n)}_B \\
0&\xi^{(n)}_B\xi^{(n)}_D&0&\xi^{(n)}_B\xi^{(n)}_D&\xi^{(n)}_B\xi^{(n)}_D\\
0&\xi^{(n)}_C\xi^{(n)}_D&0&\xi^{(n)}_C\xi^{(n)}_D&\xi^{(n)}_C\xi^{(n)}_D
\end{array} \right)
\end{equation}
with the initial condition (taking the zero order coupling $J=1$)
 ${\bf{j}}^{(0)}=(1,1,1,1,1)$.

\subsubsection{\label{sec:level3}F. Averaged values of renormalized
couplings}

 After one inflation, the new tiling has the same geometry, with
 the same relative  frequencies of vertices as the old tiling,
 however, the new
 onsite spins ${\bf{S^{(1)}}}$ and intersite
couplings ${\bf{j}}^{(1)}$ are no longer uniform from site to
site. To proceed, we define averaged quantities -- averaged
renormalization factors ${\xi^{(1)}_i}$ and averaged couplings,
for each of the seven types of site. The average couplings are
easily found, using the local environments listed for each of the
seven families in Tables 2 and 3. The simplest situation occurs
for $A$ sites, which have eight A-F links surrounding them, so
that the average coupling is just $\overline{\jmath}^{(n)}_{A} =
 j^{(n)}_{\alpha
F}$. For the six remaining sites we can similarly define averaged
couplings that are linear combinations of the $j^{(n)}$. Dropping
the superscripts, we thus have seven averaged couplings as
follows:

\bea \overline{\jmath}_{A} & =&
 j_{\alpha
F} \nonumber \\
\qquad \overline{\jmath}_{B}  &=&  (5 j_{\alpha F} +
2 j_{\alpha E})/7 \nonumber \\
\overline{\jmath}_{C}  & =&  (2 j_{\alpha F} + 3 j_{\alpha E})/5
\nonumber \\
\overline{\jmath}_{D_1}  &=& (2 j_{\alpha E} +
 j_{DD})/3  \nonumber \\
 \overline{\jmath}_{D_2}  &=&  (2 j_{\alpha E} + 2 j_{DD} +
 2 j_{D_2F})/5 \nonumber \\
\qquad \overline{\jmath}_{E}   &=& (2 j_{\alpha E}  +
 2 j_{EF})/4\nonumber \\
 \qquad
\overline{\jmath}_{F}  &=&  (j_{\alpha F} + 2 j_{EF})/3
 \eea

 Average renormalization factors ${\xi^{(1)}_i}$ are analogously
determined for each of the seven sites, and used to obtain the new
matrix $M^{(1)}$. This process is repeated and the result is a set
of recurrence relations ${\bf{j}}^{(n+1)} = M^{(n)}
{\bf{j}}^{(n)}$ with $M$ having the same structure as in Eq.8. One
can now study the evolution of the
 matrix $M$ under successive inflations.  The
 maximum eigenvalue of $M$,  $\gamma_5 \approx 0.15$.
 This results in a power law decay of the couplings for large
$n$, since ${\bf{j}}^{(n)} \approx \gamma_5 {\bf{j}}^{(n-1)}$. The
corresponding eigenvector $\vert v_5\rangle$ determines the fixed
point relative couplings.

\subsubsection{\label{sec:level3} G. Hamiltonian of inflated system}

The effective Hamiltonian after a single inflation is now written
down much as for the case of the square lattice. After the first
renormalization there are block spins at each of the
$\alpha$-class sites, whose ground state zero order energies are
$\epsilon^{(0)}_i$  and having new interblock links
${\bf{j}}^{(1)}$. $H^{(1)}( N^{(1)},\{S^{(1)}_i\},\{{\bf
j}^{(1)}\})$, where $H^{(1)}$ has the same form as the original
Hamiltonian in Eq.2 and $N^{(1)}=\lambda^{-2}N$. The original
Hamiltonian is thus decomposed into a set of independent cluster
energies and a set of intercluster terms as follows:
\begin{equation}
H = \sum_{j\in\alpha}f_j\epsilon^{(0)}_j + H^{(1)}
\end{equation}
where $j$ can take on the values A,B,C or $D_1$. The first term is
a sum over the energies of Heisenberg stars defined on the four
types of blocks $\alpha$, given by Eq.\ref{hs} or equivalently by
Eq.\ref{genhs} with $\epsilon^{(0)}_j \equiv
\epsilon(1,z,{n_0=1,\sum n_i=z})$.

\section{\label{sec:level1}V. Results }
We will discuss the calculation of the local order parameters and
then that of the ground state energy.

\subsection{\label{sec:level2}1. Local staggered magnetic moments}

The  QMC data in \cite{wess} give values of
 local order parameters. These can be defined in terms of the
 local energies around a site $i$
\begin{equation}
e_i = \frac{1}{2} J\sum_{\delta} \langle \vec{S}_i.
\vec{S}_{i+\delta}\rangle,
\end{equation}
 where the sum is over all the nearest neighbors of a given site $i$,
and the spin correlations were evaluated in the ground state. We
have added a factor $\frac{1}{2}$ per bond (that is, the bond
energy is shared equally between the two sites at each end).
 The local order parameters are defined by
\cite{note3}
\begin{equation}
m_{loc,s}^{num} = \sqrt{e_i/z}
\end{equation}
It is the quantity $e_i$ that we now wish to calculate.

 The inflation symmetry
of the quasiperiodic system allows us to define clusters on length
scales that increase as powers of $\lambda^2$. We would like a
relation between the local energies $e_i$ and the cluster
energies, denoted $E^{(n)}(z)$, evaluated as a function of $z$ for
bigger and bigger cluster size as $n$ increases. The energy {\it
per site} for a cluster of the $i$th type tends to a certain value
in the infinite size limit. We propose that this limiting value
coincides with the local energies calculated by the QMC. This is
based on the expectation that there is a fixed point distribution
for cluster energies, like the one found for the block spins, and
for the averaged couplings.

The number of terms contributing to the cluster energy is governed
by the largest eigenvalue of the block proliferation matrix $P$,
so that $E^{(n)}/7^n$ tends to a limit as $n\rightarrow \infty$.
It is this quantity that corresponds to the numerically evaluated
local energies. With this assumption, the local order parameters
at every stage of RG are found from
\begin{equation}
 m_{loc,s}^{(n)} = \sqrt{\frac{E^{(n)}(z)}{7^n z}}.
\end{equation}

We now describe how to calculate the cluster energies at each
stage of RG.

\subsubsection{Zeroth order calculation}
 The zeroth approximation was obtained in \cite{wess},
  the
energies of the clusters at this order being easily calculated
using Eq.\ref{hs} for each of the values of z, $e^{(0)} =
\epsilon^{(0)}$. The values obtained are

\begin{equation}
\{e^{(0)}_A, ... ,  e^{(0)}_F\} = \{ \frac{5}{2}, \frac{9}{4}, 2,
\frac{7}{4}, \frac{7}{4}, \frac{3}{2}, \frac{5}{4} \}
\end{equation}

 The
staggered moments corresponding to these energies are a simple
function of $z$

\bea
 m_{loc,s}^{(0)}(z)&=& \sqrt{\epsilon^{(0)}(z)/z} \nonumber \\
& = &\sqrt{\frac{z+2}{4z}} \eea

This function is plotted in Fig.\ref{third.fig}a (dashed line). In
accord with the qualitative trend of the QMC data, it shows that
$m_{loc,s}$ decreases with increasing $z$. With each additional
bond, the central spin enters into a resonant state with more and
more neighboring spins, with the result that for each individual
bond there is less amplitude for formation of a singlet.

\subsubsection{A. First order calculation}

The seven averaged couplings at this order have the numerical
values

\begin{equation}
 \{
\overline{\jmath}_{A}, ... ,\overline{\jmath}_{F} \} = \{0.14,
0.13, 0.12,0.10,0.16,0.24,0.29\} \end{equation}

These averaged couplings are used in the calculation of the ground
state energy at each of the new clusters.
 This is done using
 Eq.\ref{genhs}, along with the block spin values for the center and three
surrounding blocks deduced from Eq.5. The first order Heisenberg
star energies for each of the seven types of site are thus

\bea\left(\begin{array}{c}
\epsilon^{(1)}_A\\\epsilon^{(1)}_B\\
\epsilon^{(1)}_C\\\epsilon^{(1)}_{D1}\\\epsilon^{(1)}_{D2}\\
\epsilon^{(1)}_E\\\epsilon^{(1)}_F
\end{array} \right) =
\left(\begin{array}{c}
\epsilon(\overline{\jmath}_{A},8,\{n_0=7,\Sigma
n_i=16\})\\
\epsilon(\overline{\jmath}_{B},7,\{n_0=7,\Sigma
n_i=18\})\\
\epsilon(\overline{\jmath}_{C},6,\{n_0=7,\Sigma
n_i=16\})\\
\epsilon(\overline{\jmath}_{D1},5,\{n_0=7,\Sigma n_i=14\})\\
\epsilon(\overline{\jmath}_{D2},5,\{n_0=6,\Sigma n_i=19\})\\
\epsilon(\overline{\jmath}_{E},4,\{n_0=4,\Sigma n_i=18\})\\
\epsilon(\overline{\jmath}_{F},3,\{n_0=2,\Sigma n_i=15\})
\end{array} \right)
\eea


The energy of a cluster at first order, denoted $E^{( 1)}$,
includes this Heisenberg star energy and all zero order diagonal
terms of the sites belonging to the cluster. These first order
energies of the clusters can be expressed as follows:
 \be E^{(1)}_i = \epsilon^{(1)}_i + \epsilon^{(0)}_{anc(i)} + \frac{1}{2}
  \sum_{j=1}^z
\epsilon^{(0)}_{anc(j)}  \ee where $j=1,..,z$ are the nearest
neighbor sites of $i$, and $anc(i)$ denotes the ancestor of site
$i$. This definition takes into account the first order star
cluster energy for the cluster $i$ plus the zero energy term for
the center site, plus one-half the zero energy terms for the
surrounding sites.

 To illustrate with an example:
consider an A site on the inflated tiling, with eight nearest
neighbor F sites around it. The zero order energy term for an A
site is the block spin energy of its ancestor A site, namely,
$\epsilon^{(0)}_{A}$. The zero order energy term for F sites is
the energy of their ancestor $D_1$ block spins,
 $\epsilon^{(0)}_{D1}$. Finally, the Heisenberg Star energy for
 the A site, and with the first order effective coupling
 $\overline{\jmath}_{A}^{(1)}$
 is $\epsilon^{(1)}_{A}$.

 Consider another example
 of an F-site which has three neighbors, say
 an A site
and two E sites. The F site arises from a $D_1$ site. The zero
order block energy associated with it is therefore
$\epsilon^{(0)}_{D1}$. Similarly, the ancestors of the three
neighbors are
 an A and two C sites. They contribute
half their block energies, respectively  $\epsilon^{(0)}_{A}$ and
 $\epsilon^{(0)}_{C}$, to the total F-cluster
energy.
 The total energy of the F-cluster is found by adding four zeroth order
 terms plus
 the HS energy for F sites, which have a
 first-order coupling $\overline{\jmath}_{F}^{(1)}$.
  Other cluster energies can
be similarly obtained, and are listed below.
 \bea
E^{(1)}_A& = &\epsilon^{(1)}_A+ \epsilon^{(0)}_A + \frac{1}{2}(8
\epsilon^{(0)}_{D1})
\nonumber \\
E^{(1)}_B& = &\epsilon^{(1)}_B+ \epsilon^{(0)}_A +
\frac{1}{2}(2\epsilon^{(0)}_C +5 \epsilon^{(0)}_{D1})\nonumber\\
E^{(1)}_C& = &\epsilon^{(1)}_C+ \epsilon^{(0)}_A + \frac{1}{2}(
2\epsilon^{(0)}_{D1} + 4
\epsilon^{(0)}_C) \nonumber\\
E^{(1)}_{D1}& = &\epsilon^{(1)}_{D1}+ \epsilon^{(0)}_A +
\frac{1}{2}(\epsilon^{(0)}_B +
 4 \epsilon^{(0)}_C)\nonumber \\
 E^{(1)}_{D2}& =& \epsilon^{(1)}_{D2} +
\epsilon^{(0)}_B + \frac{1}{2}(\epsilon^{(0)}_A + 2
\epsilon^{(0)}_{D1} +
2 \epsilon^{(0)}_C) \nonumber \\
E^{(1)}_E& = &\epsilon^{(1)}_E + \epsilon^{(0)}_C +
\frac{1}{2}(\epsilon^{(0)}_A + \epsilon^{(0)}_B +
2 \epsilon^{(0)}_{D1}) \nonumber \\
 E^{(1)}_F& =& \epsilon^{(1)}_F +
\epsilon^{(0)}_{D1} + \frac{1}{2}(\epsilon^{(0)}_A + 2
\epsilon^{(0)}_C) \label{e1}
 \eea

\subsubsection{B. Second order calculation and higher orders}
 For $n=2$, the energies of
the seven clusters for the twice-inflated tiling can be written
out in terms of the energies $\epsilon^{(k)}(z)$ ($k=0,1,2$). It
is easy to obtain the explicit expressions since it suffices to
increase all the superscripts in Eq.\ref{e1} by one (so for
example the $\epsilon^{(1)}_i$ become $\epsilon^{(2)}_i$). The
zero order energy terms are also easily obtained from the
preceding order zero energy terms by use of the proliferation
matrix P defined in Eq.\ref{prolif}. We give the F cluster energy
to this order, as an example: \bea E^{(2)}_F = \epsilon^{(2)}_F +
\epsilon^{(1)}_{D1} + \frac{1}{2}(\epsilon^{(1)}_A +
2 \epsilon^{(1)}_C) + \nonumber \\
(\frac{5}{2}\epsilon^{(0)}_A +
 \epsilon^{(0)}_B +
8 \epsilon^{(0)}_C + 6 \epsilon^{(0)}_{D1}) \eea

 At third order, proceeding similarly,
 there will be a term in $\epsilon^{(3)}_F$, four
terms in $\epsilon^{(2)}$, and a certain number of terms in
$\epsilon^{(1)}$ and $\epsilon^{(0)}$. The number of blocks of
each type can be found easily using the proliferation matrix to
determine the number of ancestors of each type of block.
 In Fig.\ref{third.fig}a we have compared the $m_{s}$ obtained after zero (the
 dashed curve) with the results at
  one and two  RG steps (open circles and squares).
  After the second step, the values of
$m_{s}$ converge quickly as can be seen in Fig.\ref{third.fig}b
which shows the third (circles) and fourth order (squares) results
along with the QMC data, $m_{loc,s}^{(num)}$.

\subsubsection{C. Predictions for the full octagonal tiling}

\begin{figure}[h]
\begin{center}
\includegraphics[scale=0.850]{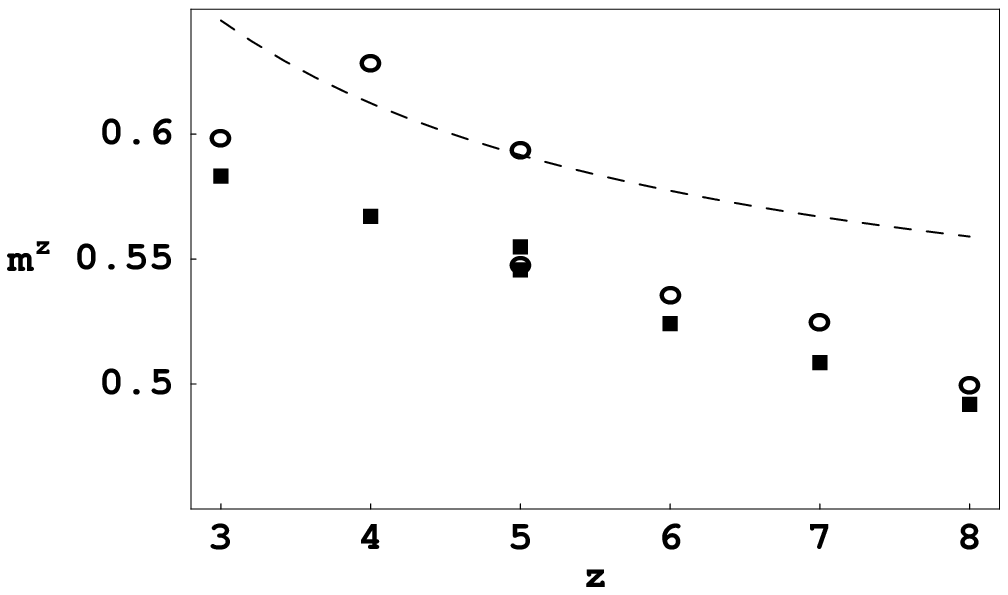}
\includegraphics[scale=0.850]{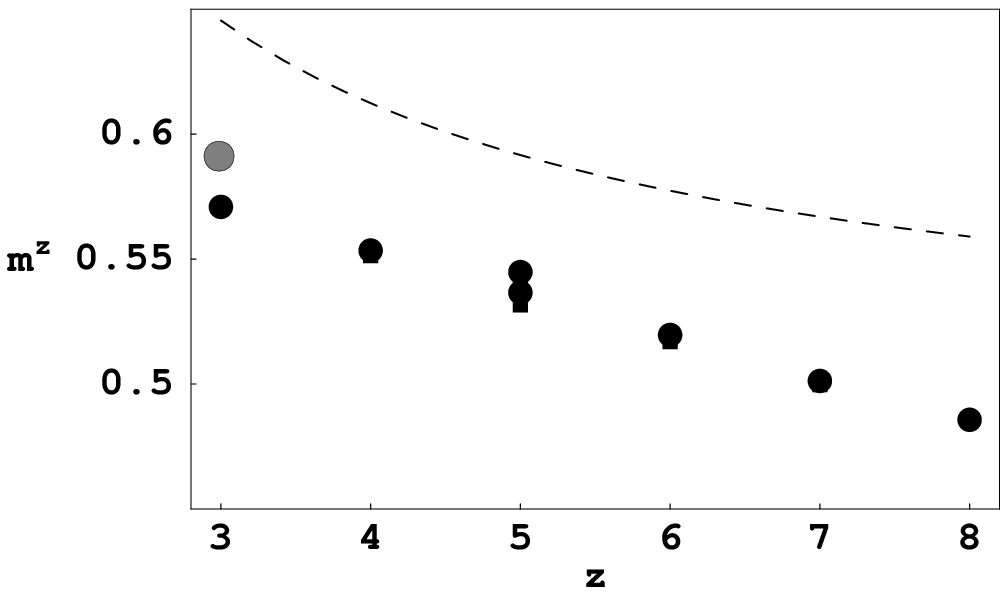}
\includegraphics[scale=0.850]{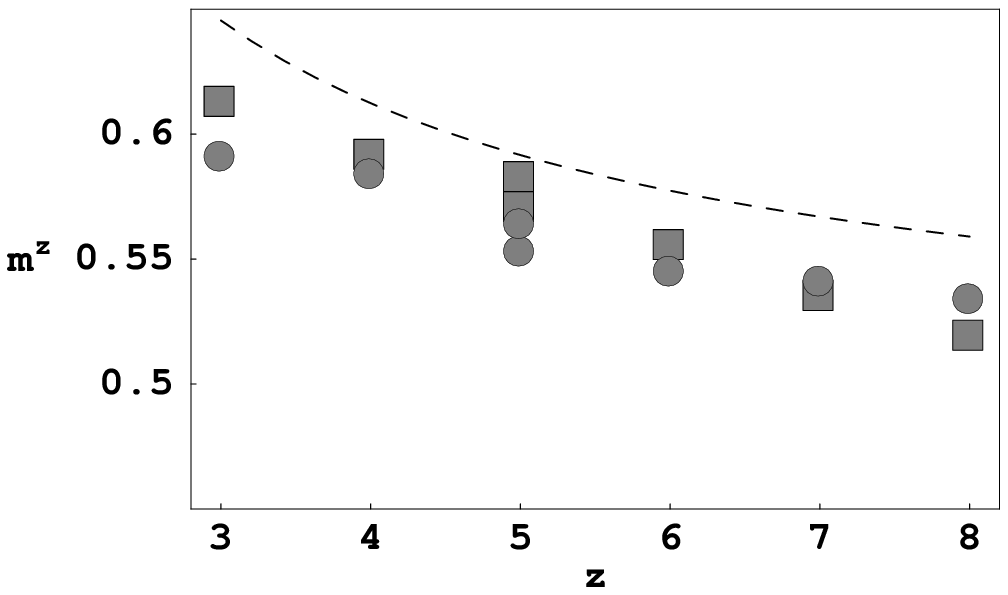}
\vspace{.2cm} \caption{ $m_s(z)$ values versus $z$ obtained for
increasing orders of RG. The zero order analytical curve is
indicated by a dashed line in each figure.
 (a)  1st (circles)and  2nd (rectangles) order RG.
 (b)  3rd (circles) and 4th (rectangles) order
 RG and QMC data for the full octagonal tiling.
 (c)  Adjusted 4th order data (grey rectangles) and
  QMC data (grey circles).
}
\label{third.fig}
\end{center}
\end{figure}

The limiting values of $m_{loc,s}$ are clearly below the QMC data.
This is to be expected, due to the bond dilution. One has to
correct for the effect of the appreciable bond dilution occurring
at C and D sites in order to obtain an estimate of the energy of
the undiluted octagonal tiling. On the one hand, the bond dilution
leads to having fewer energy terms in the Hamiltonian and
consequently underestimating the cluster energies.
 On the other
 hand, the loss of bonds is partly offset by the fact that the
dilution tends also to suppress frustration and raise the local
order parameter. An ad-hoc way to put back the ``missing
bond-energies" is to add in $half$ of the missing link energies at
each of the C and D sites. This is easily done here by adjusting
the $\tilde{z}$ values at each of the sites, $\tilde{z}_C$ goes up
from 5 to 5.5 while $\tilde{z}_{D1}$ is increased from 3 to 4.
Using this ad-hoc procedure we can get estimates for $m_{s}$
values on the original octagonal tiling. The grey squares of
Fig.\ref{third.fig}c were obtained by adjusting the $n=4$ data in
this way. As the figure shows, this procedure yields a fairly good
agreement with the QMC data. The same procedure is used to obtain
the ground state energy estimate of the full octagonal tiling in
the next section.

\subsection{\label{sec:level2}2. Ground state energy}

The ground state energy $E_{0}$ is the sum over all blocks at all
orders, of the block energies. At zero order the number of blocks
of $z$-spins $N^{(0)}(z) = N f_i$ ( i.e. proportional to the
original frequencies of occurrence given in Eq.1). The density of
vertices decreases with each inflation as $\lambda^2$, so that

\bea E_{0}/N = \sum_{i \in \alpha} f_i(\epsilon^{(0)}_i +
\frac{1}{\lambda^{2}}\epsilon^{(1)}_i + ... \frac{1}{\lambda^{2n}}
\epsilon^{(n)}_i + .....) \label{gse}\eea

 The block energies $\epsilon^{(n)}$ are the energies of blocks with a spin
$S_0^{(n)}$ at the center, with effective couplings
${\overline{\jmath}}^{(n)}$ to the $S_i^{(n)}$ surrounding spins.
 The series
for the energy gives $e_{0} \approx -0.51$. We can estimate the
effect of bond dilution, as was done for the local order
parameters. Using the corrected values of $\tilde{z}$ explained in
the last section, one finds an adjusted ground state energy of
about $-0.59$. This value of the GS energy is significantly
smaller in absolute value than the value deduced from the QMC data
in \cite{wess}. We recall that this was true of the square lattice
calculation as well. In that case, the RG calculation of Delgado
and Sierra
 was already noted in
\cite{sierra} to underestimate the bonding energies of pairs of
spins because of the inadequacy of first order perturbation theory
around the Neel state. The same is presumably true of our RG on
the octagonal tiling. For the former case the RG calculation was
compared with the terms of a $1/S$ expansion of the ground state
energy, and shown to lack the subleading order term, resulting in
the observed discrepancy of values.

 On the square lattice, $e_{0}$ has been determined numerically
\cite{sand} to high precision
 to be $-0.6694$, while finite size scaling for the tiling \cite{wessnew}
  obtains a value of $-0.6581$.
  The closeness of the values
 obtained for these two very different problems is rather
 surprising. It is a probable that this close proximity of values
 is due to the fact that
 the octagonal tiling, with
its two sublattice structure and its average coordination number
of 4. The differences must arise from the next nearest neighbor
distributions which differ for the two systems, although this
remains to be verified by explicit calculation.

\section{\label{sec:level1}VI. Discussion and conclusions}

In conclusion, we have presented an approximate RG scheme for
ground state properties of a two-dimensional quasiperiodic tiling
that can be solved after bond dilution. Other approximations
involve the truncation of the number of distinct sites and the
number of distinct links, and replacing local couplings around
sites by average values in order to simplify the effective
Hamiltonian after every inflation. The results obtained for the
diluted tiling were used to get estimates for the undiluted
tiling. Despite these approximations, we believe the model solved
is close to the perfect two dimensional quasiperiodic structure,
and it allows for a rather detailed solution of real space
properties of these heirarchical structures. The results obtained
by RG for local order parameters
 are close to those calculated for
the full undiluted model, after our adjustment procedure. It thus
appears that the model takes into account the most relevant
aspects of
 the quasiperiodic geometry of the octagonal tiling.

 The RG method presented is
less good at obtaining the ground state energy, similar to the
situation already noted for the square lattice by Sierra and
Martin-Delgado, who showed that a better result is obtained by
going to second order of perturbation theory to obtain the
effective Hamiltonian after renormalization. Concerning the
proximity of values
 of the ground state energy in these two systems, our calculation
 is not accurate enough to explain this observation.
 A calculation to higher order
 would involve further nearest neighbor sites, improve
 the energy estimate and perhaps help to explain the
 small energy difference between the tiling and the
 square lattice. It would be interesting as well to compare
 results for
 other bipartite two dimensional tilings, including the Penrose tiling.

 The zero temperature magnetic state of this quasiperiodic
 Heisenberg antiferromagnet has a structure factor with peaks that
 can be indexed using the four dimensional indexing scheme (see
 Appendix). The positions of the peaks is very simply related to
 the positions of the peaks of the paramagnetic state: they are
 situated halfway in between. In other words, the paramagnet is
 indexed by four integers, while the antiferromagnet has
 half-integer entries, corresponding to the antiferromagnetic
 vector ${\bf q} =
 \{\frac{1}{2},\frac{1}{2},\frac{1}{2},\frac{1}{2}\}$. This is the
 quasiperiodic
 analogue of the square lattice where just such a shift occurs in
 reciprocal space and corresponds to the antiferromagnetic
 vector ${\bf q} =
 \{\frac{1}{2},\frac{1}{2}\}$ (see \cite{ron} for a discussion along
 with a simple one dimensional version of a quasiperiodic
 antiferromagnet). The real life quasiperiodic compound
 ZnMgHo was studied by neutron scattering and shown to have short
 range antiferromagnetic correlations below about 20K. These
 correlations lead to a magnetic superstructure that is, as for
 our two dimensional model, shifted with respect to the
 paramagnetic state. The antiferromagnetic vector that best fits
 the data has a more complicated value than the simplest form for
 a 3d quasiperiodic antiferromagnet ($q_i = \frac{1}{2}, i =
 1,6$). This is because the magnetic unit cell is much larger for
 the three-composant system, due to the fact that only the Ho
 sites carry a magnetic moment, resulting in smaller spacings between
 peaks in reciprocal space.

Finally, the RG scheme presented here can be adapted to discuss
other discrete quasiperiodic models, such as tight-binding models
for electrons hopping between vertices of the tiling. It should
provide a useful theoretical framework for describing
quasiperiodic tilings in general.

\appendix
\section{Appendix. The cut-and-project method.}

\subsection{\label{sec:level1} 1. One dimensional example}
The cut-and-project method of obtaining quasiperiodic tilings is
easiest to illustrate in the case of the celebrated
one-dimensional tiling -- the Fibonacci chain(see Luck's review in
\cite{lucrev}). The Fibonacci chain comprises two basic tiles or
line segments of two different lengths, "long"(L) and "short"(S)
arranged in a deterministic sequence. The Fibonacci chain can be
generated iteratively from a single S segment using the following
substitution rules: replace each S by an L, and each L by SL. Two
successive segments of the infinite chain are shown below to
illustrate the substitution rules (dashed lines represent L, thick
lines S)

\begin{figure}[ht]
\begin{center}
\includegraphics[scale=0.60]{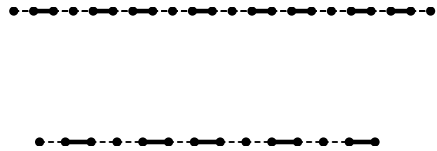}
\vspace{.2cm} \caption{two segments of the Fibonacci chain }
\label{four.fig}
\end{center}
\end{figure}

\begin{figure}[ht]
\begin{center}
\includegraphics[scale=0.60]{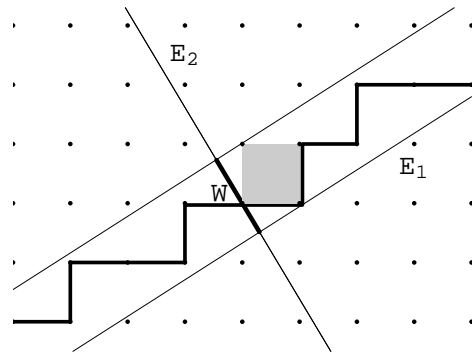}
\vspace{.2cm} \caption{ Cut-and-project method. Selected edges of
a square lattice are projected onto the parallel space ($E_{1}$).
The edges that are selected have perpendicular space ($E_{2}$)
projections that fall within the segment marked W.
 } \label{proj.fig}
\end{center}
\end{figure}

The Fibonacci sequence of segments, or tiles, can be generated by
projecting selected edges of a two-dimensional square lattice onto
the one dimensional "physical space" $E_{1}$ as shown in
Fig.\ref{proj.fig}. The vertical and horizontal edges project onto
the S and the L tiles respectively. The
 orientation of $E_{1}$ is given by $\tan^{-1} 1/\tau$ (where $\tau = (\sqrt{5}
+1)/2$ is the golden mean, a solution of $\tau^2-\tau-1=0$), an
irrational slope, so the tile sequence never repeats. The edges
selected for projection onto $E_{1}$ obey the following condition:
the projection of the edge onto the perpendicular space $E_{2}$
 must fall within the "window of
selection" W (indicated by the thick line segment representing the
projection of the unit square shown in grey). A finite sequence of
twelve edges satisfying this condition are shown in bold in
Fig.\ref{proj.fig} and they result in the projected structure
...LSLLSLLSLSLL...

 \subsection{\label{sec:level1} 2. Two dimensional case}
  In analogy with the one dimensional case, the
octagonal tiling is obtained from the projection onto
$E_{\parallel}$ (the physical two-dimensional space) of a subset
of vertices of a four-dimensional cubic lattice. The subspaces
$E_{1}$,$E_{2}$ are now two-dimensional, and are invariant under
eight-fold rotations in the four dimensional space. The
orientation of the physical plane $E_{\parallel}$ is given by the
number $\lambda = 1+ \sqrt{2}$, one of the solutions of $\lambda^2
- 2 \lambda -1 =0$. The tiles, which are projections in this plane
of the 8 faces of the 4d cube, are squares and $45^o$ rhombuses.
The vertices (and edges) that are selected for projection in the
two dimensional
 perpendicular space $E_{2}$ must
 fall within the window of selection shown in Fig.\ref{perpspa.fig}.
 This octagon-shaped
area is delimited by the projection of the sides of a four
dimensional unit cube.  The octagonal tiling has by construction
the eight-fold symmetry in the weak sense already described. There
are six different kinds of nearest neighbor configurations for
vertices of the tiling, denoted A through F as shown in Fig.2.

\subsubsection {A. Domains of acceptance}
\begin{figure}[ht]
\begin{center}
\includegraphics[scale=0.50]{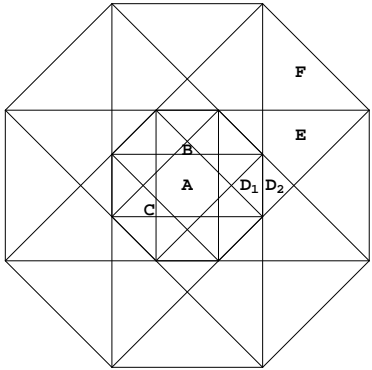}
\vspace{.2cm} \caption{ Projection into perpendicular space of
vertices of the octagonal tiling. Domains corresponding to the six
families are labeled ( eight-fold symmetry determines labels of
unmarked domains) } \label{perpspa.fig}
\end{center}
\end{figure}

 The maximum coordination
number $z=8$ corresponds to the eight-fold symmetric A sites.
These sites
 possess perpendicular space projections that always fall within the
central octagon labeled A in Fig.\ref{perpspa.fig}. Similarly, the
projection of all $z=7$ sites falls within one of the eight
triangular regions adjoining the central octagonal domain. The
remaining $z$ values likewise correspond to the domains labeled
accordingly in Fig.\ref{perpspa.fig}, which has exact eight-fold
symmetry. The ratio of side lengths of similar polygons are either
$\lambda$, or $\lambda^2$.


\vskip 0.5cm

\subsubsection {B. Inflations and deflations}

 For the octagonal
tiling, the inflation transformation is given by a $4\times 4$
matrix acting in the four dimensional cubic lattice, satisfying
$U^2 - 2 U -1 =0$ and having entries of 0 or 1 only. Projecting
the subset of points selected by $u$ leads to a bigger tiling of
the same type as the original one. Only the highest $z$ sites
remain selected, while the others disappear. The sites that remain
are those within the middle octagon, the $\alpha$ family: A, B, C
and $D_1$. The sites that disappear correspond to the region
outside the middle octagon. The perpendicular space representation
also allows to determine rapidly the new $z$ values of the sites
that remain: one simply redraws the acceptance domains of
Fig.\ref{perpspa.fig} after rescaling, inside the middle octagon.
Thus a point that was previously in the $D_1$ domain will find
itself in the F domain, and the C domains
 map into E domains. A sites remain A sites if
they are close to the center of the diagram, otherwise they become
one of the other $\alpha$ sites after inflation. The four
categories of A sites mentioned in section II.3 differ by their
distance from the origin in perpendicular space. The perpendicular
space projection of a site determines its evolution under
inflation -- the closer a site is to the center of the octagonal
selection window, the longer it remains an A site under successive
inflations. Also, one clearly sees that D-sites come in two types,
with different perpendicular space domains. The following table
resumes the old and new site types after inflation:

\begin{tabular}{lccccccr}
A& $\rightarrow$&A&or&B&or&C&or $D_1$ \\
B &$\rightarrow$& $D_2$&&&&& \\
C &$\rightarrow$& E &&&&& \\
$D_1$ &$\rightarrow$& F&&&&& \\
\end{tabular}

 The number of sites per unit area is
reduced by the scale factor $\lambda^2$ after each inflation, and
the relative frequencies of occurrences of each of the seven
families of sites is invariant. The frequency of occurrence of the
$i$th family is proportional to the area occupied by that family
in the perpendicular space projection. It can thus be easily
verified using Fig.\ref{perpspa.fig} that these frequencies are: $
f_A=\lambda^{-4};
 f_B=\lambda^{-5}; f_C=2\lambda^{-4};
f_{D1}=\lambda^{-3}= ; f_E = 2\lambda^{-2}; f_F = \lambda^{-2}$.
The average coordination number is exactly 4, as can be checked
using the frequencies given. The interested reader can find these
and other important geometric and algebraic properties of the
system described for example in \cite{dunref}.

\subsubsection {C. Reciprocal space and structure factor}
 The
diffraction peaks of the octagonal tiling are found at positions
given by projections into 2d of reciprocal lattice vectors ${{\bf
a}_i}$ of the 4d cubic structure. The intensities of the peaks are
not uniform however, but depend on the Fourier Transform (FT) of
the finite selection window. The main features of the diffraction
pattern are thus (see Belin et al in \cite{belin} for more on the
topic)
\begin{itemize}
\item The peaks have an eight-fold symmetry around the peak at the
origin. \item peaks occur at positions corresponding to the set of
integers ${h,k,m,n}$ representing the projection into the 2d plane
of the 4d vector ${\bf q} = h {\bf a}_1 + k {\bf a}_2 + m {\bf
a}_3 + n {\bf a}_4$. That is, one can index peaks by a set of four
integers. \item Intensities are highly dependent on the value of
${\bf q}$ since the FT of the selction window is oscillatory and
long ranged. The set of eight most intense peaks nearest the
origin is used to define a quasi Brillouin zone for the tiling.
\end{itemize}

\subsubsection {D. Approximants and some of their properties}
Numerical studies
of quasiperiodic systems are performed on finite pieces of the
infinite system. In particular, it has been pointed out that
periodic boundary conditions are preferable to open or closed
boundary conditions in terms of eliminating spurious states and
eigenvalues. A periodic approximants is a structure that can be
periodically continued and can be augmented in size so as approach
arbitrarily close to the perfect infinite structure.

 This is,
again, easiest illustrated by going back to the Fibonacci chain.
In the cut and project technique, it should be clear that if one
tilts the irrationally oriented selection strip away from the
special angle, one will obtain a periodically repeating chain
every time the slope is rational.

 $\tau^{-1}$ has a series of
approximants given in terms of the Fibonacci numbers as
follows:$\{\alpha_1,...\}$ =
$\{1,\frac{1}{2},\frac{2}{3},\frac{3}{5},...,\frac{F_k}{F_{k+1}}...\}$,
where $F_k$ is the $k$th term in  the Fibonacci sequence defined
by the recurrence relation

\bea F_{k+1} = F_{k}+F_{k-1} \nonumber \\
F_0 = F_1 = 1 \eea

with $F_0=F_1=1$.
  By increasing the value of the
denominator of the rational number $F_n/F_{n+1}$ -- i.e. by
choosing increasingly longer approximants of the golden mean --
one will get a structure of period $F_{n+2}$. The finite sequences
of L and S within the approximant are the same as those found in
an infinitely long chain.

For the two dimensional case, Ref.\cite{dunmoss} describes how to
obtain square approximants to the octagonal tiling by the
projection method. These are obtained from the approximants to the
silver mean which depend on ratios of the so-called Octonacci
sequence : $\lambda_k = O_{k+1}/O_k$ where

\bea O_{k+1} = 2O_{k}+ O_{k-1} \nonumber \\
O_1 = 1; \qquad O_2 = 2 \eea

with $O_1=1;O_2=2$. These are the finite size systems used for a
number of numerical studies including the quantum Monte Carlo
calculations. The first few square approximants have the following
sizes

\begin{center}
\begin{tabular}{|l|c|c|c|r|}
\hline
k &\quad 2 \quad & \quad 3 \quad & \quad 4 \quad & \quad 5 \quad\\
 $N_k$&239 & 1393 &8119 & 47321 \\
\hline
\end{tabular}
\vskip 0.5cm \small{Table A.1. Number of sites in the first four
square approximants}
\end{center}

We will list some features of these approximants that may be
important to bear in mind depending on the models studied.

\begin{itemize}
\item Reflection symmetry (exact) with respect to the bottom
left-top right diagonal

\item $90^o$ Rotation symmetry around the center (approximate)

\item Odd parity of repetition. By this is meant that one changes
sublattice when one goes from a site to its first periodic
repetition along either x or y directions. For a number of
numerical calculations it is easiest to restore the bipartite
property by taking a system size doubled along both directions
(i.e. quadrupled unit cell with respect to the sizes given in the
Table).

\item Inflation relation between approximants. Fig.\ref{appen.fig}
shows a small approximant superimposed on the next largest one.

\begin{figure}[ht]
\begin{center}
\includegraphics[scale=0.40]{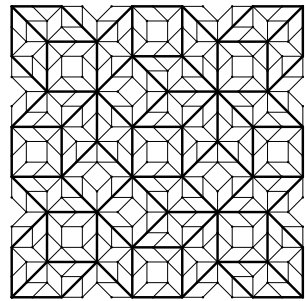}
\vspace{.2cm} \caption{Superposition of two successive
approximants} \label{appen.fig}
\end{center}
\end{figure}

\end{itemize}


\begin{references}
\bibitem{note2} The two
kinds of D-site also can be distinguished in real space by
counting the number of second nearest neighbors.
\bibitem{note3} Note a slight change of definition of $m_{s}$
from that of ref.3.
\end{references}
\end{document}